\documentstyle[aps,tighten,preprint,eqsecnum,graphicx]{revtex}
\begin{document}
\draft
\title{Finite-size scaling investigations in the quantum
$\varphi^4$-model with long-range interaction}
\author{Hassan Chamati and Nicholay S. Tonchev}
\address{Institute of Solid State Physics,
72 Tzarigradsko chauss\'ee, 1784 Sofia, Bulgaria}

\maketitle
\begin{abstract}
In this paper, we study in details the critical behavior of the ${\cal
O}(n)$ quantum $\varphi^4$ model with long-range interaction decaying
with the distances $r$ by a power law as $r^{-d-\sigma}$ in the large
$n$-limit. The zero-temperature critical behavior is discussed. Its
alteration by the finite temperature and/or finite sizes in the space
is studied. The scaling behaviours are studied in different regimes
depending upon whether the finite temperature or the finite sizes of
the system is leading. A number of results for the correlation length,
critical amplitudes and the finite-size shift, for different
dimensionalities between the lower $d_<=\sigma/2$ and the upper
$d_>=3\sigma/2$ critical dimensions, are calculated.
\end{abstract}

\section{Introduction}\label{introduction}
The vast majority of existing analytical and numerical work on
finite-size effects has been in regimes where the quantum effects are
entirely ruled out~\cite{binder92}. In this case both the statics and
the dynamics can be described by classical statistical models. At the
critical point the bulk thermodynamic functions of these models are
singular. For these systems the finite-size scaling theory asserts
that the singularities holding at the thermodynamic limit are altered
depending upon the nature of the geometry to which the system is
confined and the imposed boundary
conditions~\cite{barber83,privman90}. Indeed various type of
geometries, can be considered, depending on the number of the finite
sizes in the model.

The ${\cal O}(n)$ vector models are extensively used to explore the
finite-size scaling theory, using various methods and techniques. For
finite $n$ the most used method is that of renormalization
group~\cite{zinnjustin96}. The most investigated case is the one
corresponding to the limit $n=\infty$ (this limit includes also the
mean spherical model)~\cite{privman90}. In this limit, these models
are exactly soluble for arbitrary dimensions and in a general
geometry. The most part of these investigation are limited to systems
in which the forces are short-range. To test the finite-size scaling
idea, when the interaction is of a long range (varying with a power
law), the only used model is the mean spherical
model~\cite{brankov92}.

In recent years there has been an increased interest in the theory of
zero temperature quantum phase transitions~\cite{sondhi97,cesare97}.
Distinctively from temperature driven critical phenomena, these phase
transitions occur at zero temperature as a function of some
non-thermal control parameter (or a competition between different
parameters describing the basic interaction of the system), and the
relevant fluctuations are of quantum rather than thermal nature. In
this type of critical phenomena the time plays a crucial and
fundamental role. The coupling of statistics and dynamics that is
inherent to quantum statistical problems introduces effective
dimensionalities in the hyperscaling laws, i.e. the space
dimensionality $d$ is replaced by $d+z$ ($z$ is the dynamic critical
exponent). In this case the inverse temperature acts as a finite size
in the ``imaginary-time'' direction for the quantum system at its
critical point. This allows the investigation of scaling laws for
quantum systems near the quantum critical point in terms of the theory
of finite-size scaling~\cite{sachdev96} or using the conformal field
theory techniques by mapping the bulk system in a finite one
\cite{christe93}. The ${\cal O}(n)$ symmetric vector models are also
used in the exploration of of the quantum critical phenomena and for
the investigation of the scaling properties of such phase transitions.
The quantization of classical ${\cal O}(n)$ is performed with the help
of the Trotter formula which maps the quantum model on a classical one
with $z$ additional effective dimensions.

In systems showing quantum critical behaviour the temperature plays
two different roles. For temperatures low enough, quantum effects are
essential. In this case the temperature affects the geometry to which
the system is confined adding a ``new'' sizes to the eucludean
space-time coordinate system. By raising the temperature, the system
is driven away from quantum criticality. At high temperatures,
however, the size in the ``imaginary-time'' direction becomes
irrelevant in comparison with all length scales in the system. In this
case we have a classical system in $d$ dimensions and the temperature
is just a coupling constant in the classical critical behaviour.

In this paper we present a detailed investigation of the scaling
properties of the quantum ${\cal O}(n)$ vector $\varphi^4$ model with
long-range interaction. Our study will include the quantum as well as
the finite-size effects and their influence on the critical behaviour
and the critical amplitudes. We will also check the influence of the
interaction range on the critical behaviour. These interactions enter
the exact expressions for the free energy only through their Fourier
transform which leading asymptotic is $U(q)\sim q^{\sigma^*}$, where
$\sigma^*=min(\sigma,2)$ \cite{fisher72}. As it was shown for bulk
systems by renormalization group arguments $\sigma\ge 2$ corresponds
to the case of finite (short) range interactions, i.e. the
universality class then does not depend on $\sigma$ \cite{fisher72}.
Values satisfying $0<\sigma<2$ correspond to long-range interactions
and the critical behaviour depends on $\sigma$. On the above reasoning
one usually considers the case $\sigma>2$ as uninteresting for
critical effects, even for the finite-size treatments \cite{fisher86}.
So, here we will consider only the case $0<\sigma\le 2$.

The paper is devised as follows: in Section~\ref{FSSQ}, we present
some predictions which extend the finite-size scaling to quantum
systems. We also discuss the interplay of quantum and finite-size
effects on the quantum critical behaviour. We comment on the
anisotropy caused by the presence of both effects. In
Section~\ref{phi4}, we review, briefly, the $\varphi^4$-model with
long-range interaction and present the saddle point equation. We
investigate the low temperature behaviour of the bulk model in
Section~\ref{tempbulk}. In Section~\ref{zerotemperature}, we
investigate in details the finite-size behaviour at zero temperature.
Section~\ref{interplay} is devoted to some comments about the low
temperature and finite-size effects of the system. In
Section~\ref{discussion} we discuss our result briefly. In the
remainder of the paper we present some details of the calculations.

\section{Finite-size scaling and quantum critical behaviour}
\label{FSSQ}
Divergent length scales play a crucial role in continuous phase
transitions. Unlike classical models, where the scaling can be done
equally for all ``spatial'' dimensions, quantum models are anisotropic
in general, and therefore the ``space'' and ``imaginary-time''
directions will not scale in the same fashion. According to general
hypothesis of finite-size scaling theory~\cite{privman84} extended
here for quantum (anisotropic) system, a physical quantity ${\cal A}
\left(r,h,L,T\right)$ (with $r$ measuring the distance from the
critical point, $h$ is an ordering field coupled to the order
parameter, $T$ is the temperature and $L$ is the size of the system),
which may be singular at the critical point in the thermodynamic
(bulk) limit at the quantum critical point ($r=0$), will scale like
\begin{equation}\label{FSSA}
{\cal A}(r,h,T,L)=b^p {\cal A}_s\left(rb^{1/\nu},
hb^{\Delta/\nu},Tb^z,bL^{-1}\right).
\end{equation}
In the scaling form~(\ref{FSSA}), $p$ corresponds to the engineering
dimension $d+z$ of the system in the case when the scaling function
refers to the singular part of the free energy, and it is the divided
by $\nu$ critical exponent measuring the divergence of the bulk
thermodynamic function ${\cal A}$ at the critical point for the other
physical quantities of interest (for the correlation length $p=1$, for
the susceptibility $p=\gamma/\nu$, etc). Depending upon the choice of
the renormalization group rescaling factor $b$ we obtain different
scaling functions ${\cal A}_s$, which are related among each other by
some appropriate change of the scaling variables.

Before starting to discuss the general form of the scaling universal
function ${\cal A}_s$, we will discuss the two limiting cases which
were subjects of several investigations during the past two decades.

The first case corresponds to zero temperature and is called hereafter
the quantum critical behaviour. Here equation~(\ref{FSSA}) reduces to
\begin{equation}\label{FSST0}
{\cal A}(r,h,0,L)=b^p {\cal A}_s^L\left(rb^{1/\nu},
hb^{\Delta/\nu},bL^{-1}\right).
\end{equation}
Following Ref.~\cite{brezin82}, we choose the rescaling factor $b$ to
be proportional to the linear size of the system. Then, we obtain
\begin{equation}\label{FSSL}
{\cal A}(r,h,0,L)=L^p {\cal A}_s^L\left(rL^{1/\nu},
hL^{\Delta/\nu}\right).
\end{equation}
Here the situation resembles that of systems exhibiting classical
(thermal) phase transition.

It is possible to get another result for the scaling function in the
right hand side of equation~(\ref{FSSL}), if one considers the
variable $\tilde r L^{1/\nu}$ instead of the first variable of the
scaling function ${\cal A}_s^L$. The parameter $\tilde r$ is
introduced in such a way to account for the shift of the critical
quantum parameter to the value corresponding to the rounding of the
thermodynamic functions, when the number of the infinite dimensions
$d'$ is less than its lower critical dimension $d_<$. For the opposite
case we have just a shift of the critical control parameter. In this
case we find that the critical exponents are those of a $d'$
dimensional system (see for example~\cite{barber83}). By definition
the finite-size shift is given by
\begin{equation}\label{shiftr}
\tilde r=r+\epsilon(L), \ \ \ \ \ \ \lim_{L\to\infty}\epsilon(L)=0.
\end{equation}
In general we have
\begin{equation}\label{shiftnu}
\epsilon(L)\sim L^{-1/\nu},
\end{equation}
and so this quantity shrinks to zero in the thermodynamic limit. When
the arguments of the scaling functions get replaced by zero, we will
obtain universal critical amplitudes, characterizing the whole class
of universality. Here we have to emphasize that the values of the
universal critical amplitudes are different depending upon the point
in which they are calculated i.e. $r$ or $\tilde r$.

The second case we consider is the one corresponding to the bulk
system ($L=\infty$) at finite-temperature. In this case the scaling
form (\ref{FSSA}) transforms into (see Ref.~\cite{cesare97} and
references there)
\begin{equation}\label{FSST}
{\cal A}(r,h,T,\infty)=T^{-p/z} {\cal A}_s^\tau\left(rT^{-1/z\nu},
hT^{-\Delta/z\nu}\right),
\end{equation}
where we used the relation $b=T^z$ between the temperature and the
rescaling parameter $b$. The same predictions for the shift and the
critical amplitudes remain valid here. We find it convenient to use
$L_\tau\sim T^{-z}$ as a linear ``temporal'' size instead of the
inverse temperature.

The general case corresponding to both quantum and finite-size effects
can be studied by considering a system with larger dimension $p=d+z$
confined to a general geometry of the form $L^{p-z}\times L_\tau^z$
\cite{korucheva92}. The fact that the inverse temperature can be used
as an additional size in the imaginary-time direction creates some
anisotropy in the system. This property will lead to the establishment
of some change in the scaling properties of the finite quantum system.
In the general case we can consider the quantum to classical and the
finite-size to the bulk system different crossover phenomena. It is
easy to convince oneself about this statement by thinking on the
dynamic critical exponent which is different for different quantum
systems. The situation of the combined investigations of finite-size
scaling and quantum to classical crossover is similar to the one
formulated in the framework of the phenomenological study of
finite-size scaling in anisotropic systems~\cite{binder89}. In this
case one can investigate the interplay of quantum and finite-size
scaling. For example in the case when the quantum effects are leading
compared to the finite-size (this case will be called here
``low-temperature'' case) i.e. $L_\tau\ll L$, from
equation~(\ref{FSSA}) to first order in $L_\tau/L$, we expect to
obtain
\begin{eqnarray}\label{correction}
{\cal A}(r,h,T,L)&=&L_\tau^p {\cal
A}_s^\tau\left(rL_\tau^{1/\nu},hL_\tau^{\Delta/\nu}\right)\nonumber\\
&&+L_\tau^{p+1}L^{-1}{\cal A}_s^{\tau
L}\left(rL_\tau^{1/\nu},hL_\tau^{\Delta/\nu}\right)
\end{eqnarray}
instead of equation~(\ref{FSST}). This shows how the finite-size
effects give rise to some corrections to the quantum scaling.
Following the same reasoning in the case of ``very low-temperature''
i.e. $L_\tau\gg L$, when the finite-size contributions to the ground
state energy and its derivatives are dominant compared to those coming
from the quantum effects, we find
\begin{eqnarray}\label{cortosca}
{\cal A} (r,h,T,L)&=&L^p {\cal
A}_s^L\left(rL_\tau^{1/\nu},hL_\tau^{\Delta/\nu}\right)\nonumber\\
&&+L_\tau^{-z} L^{p+z}{\cal A}_s^{L\tau}
\left(rL_\tau^{1/\nu},hL_\tau^{\Delta/\nu}\right),
\end{eqnarray}
showing a correction to the zero-temperature finite-size scaling due
to the temperature. These ideas has been tested in the framework of
the spherical quantum rotor model in Ref.~\cite{chamati97,chamati98}.
We will compare the scaling forms~(\ref{correction}) and
(\ref{cortosca}) with the available analytical results obtained below.

\section{The free energy and the gap equation}\label{phi4}
The quantum $\varphi^4$-model we are going to investigate its quantum
critical and finite-size scaling properties is (for a review of the
applicability of this model in exploring quantum critical phenomena
see Ref.~\cite{cesare97})
\begin{equation}\label{model}
{\cal H}\left\{\varphi\right\}=\frac12\int_0^{1/T} d\tau\int dx
\left[\left(\partial_\tau\varphi\right)^2+
\left(\nabla^{\sigma/2}\varphi\right)^2+r_0\varphi^2
+\frac {u_0}2\varphi^4\right],
\end{equation}
where $\varphi$ is a short hand notation for the space-time dependent
$n$-component field $\varphi(x,\tau)$, $u_0$ and $r_0$ are model
constants and $T$ is the temperature. In (\ref{model}) we assumed
$\hbar=k_B=1$ and the size scale is measured in units in which the
velocity of excitations $c=1$.  Here we will consider
periodic boundary conditions. This means
\begin{equation}\label{fourier}
\varphi(x,\tau)=\sqrt{\frac TV}\sum_{\bbox k,\omega_l}\varphi(\bbox k,
\omega_l)\exp\left(\imath \bbox k\cdot x-\imath\omega_l\tau\right),
\end{equation}
where $\omega_l=2\pi T l$ (with $l=0,\pm1,\pm2,\cdots$) are the
Matsubara frequencies for bosonic systems, ${\bbox k}$ is a discrete
vector with components $k_i=2\pi n_i/L$, $n_i=0,\pm1,\pm2,\cdots$ and
a cutoff $\Lambda$, and $V=L^d$ is the volume of the system. We note
that the second term in the model transforms into $|{\bbox
k}|^\sigma\varphi^2({\bbox k},\omega)$ in the momentum representation,
where the parameter $0<\sigma\leq2$ accounts for short-range and
long-range interaction as well.

The partition function of the Hamiltonian~(\ref{model}) reads
\begin{equation}\label{partition1}
{\cal Z}=\int{\cal D}\varphi\exp\left(-{\cal
H}\left\{\varphi\right\}\right).
\end{equation}
Note that in the low-temperature limit $T\ll\Lambda$ the integral over
$\tau$ in ${\cal H}\{\phi\}$ (see (\ref{model})) can be extended over
the whole temperature axis to give an effective $\varphi^4$ model in
$d+z$ dimensions with a quantum control parameter $r_0$. In the high
temperature limit $T\gg\Lambda$ the upper limit in the integral over
$\tau$ is small. This offers us the possibility to write the
Hamiltonian as a classical $\varphi^4$ model in $d$ dimensions. Now
using a standard decoupling procedure based on the
Hubbard-Stratonovich transformation in (\ref{partition1}), which
introduce an auxiliary filed $\psi$, one gets
\begin{eqnarray}\label{decoupled}
{\cal Z}&=&{\cal C}\int{\cal D}\psi{\cal D}\varphi\exp\left[-\frac12
\int_0^{1/T}d\tau\int dx
\left(\partial_\tau\varphi\right)^2\right.
+\left(\nabla^{\sigma/2}\varphi\right)^2\nonumber\\
&&+\left.r_0\varphi^2 +\psi\varphi^2-\frac1{2u_0}\psi^2\right].
\end{eqnarray}
Using the fact that the field $\varphi$ is $n$-component we decompose
the integral over $\varphi$ into an $n$-dimensional gaussian integral,
which can be performed easily, leading to
\begin{equation}\label{lastz}
{\cal Z}={\cal C}\int{\cal D}\psi \exp\left[\frac {n\beta
V}{4u_0}\psi^2-\frac{n}{2} \ {\rm Tr}
\ln[r_0+\psi+\partial_\tau^2+\nabla^\sigma]\right].
\end{equation}
In the last expression we assumed that the field $\psi$ is time and
space independent. For large $n$ we use the saddle point method to
evaluate the integral over $\psi$ in (\ref{lastz}). Finally, we obtain
the free energy per particle ${\cal F}=-\left(T/V\right) {\rm Tr}
\ln{\cal Z}$ in the momentum space
\begin{mathletters}\label{scalingpro}
\begin{equation}\label{freeenergy}
{\cal F}=-\frac{1}{4u_0}(\phi-r_0)^2+
\frac{1}{V}\sum_{\bbox k}\ln2\sinh\left[\frac{1}{2T}\sqrt{\phi+
|\bbox k|^\sigma}\right]
\end{equation}
and the saddle point equation
\begin{equation}\label{phi}
\phi=r_0+u_0\frac{T}{V}\sum_{\bbox k,m}\frac1{\phi+(2\pi mT)^2+|\bbox
k|^\sigma},
\end{equation}
\end{mathletters}
where, for convenience, we used the shifted parameter
$\phi\equiv\psi+r_0$ instead of $\psi$ itself.

Equations~(\ref{scalingpro}) are our starting point to explore the
finite-size and quantum effects on the bulk critical behaviour of the
model (\ref{model}). Let us note that in the particular case of
short-range forces $\sigma=2$, we recover the results of
Ref.~\cite{chen98} utilized to investigate the finite-size scaling
when the quantum effects are absent. When the quantum effects are
relevant, equation~(\ref{phi}) was used in Ref.~\cite{morf77} in order
explore the quantum critical behaviour and to calculate the critical
exponents.

To extract the physics from model~(\ref{model}), we calculate the
susceptibility $\chi$. In the large $n$ limit, we obtain
\begin{equation}\label{suc}
\chi\equiv\int d^dx\langle\varphi(x)\varphi(0)\rangle= \frac1\phi.
\end{equation}
This is a particular case of the general correlator $\chi({\bbox
k},\omega)=\langle\phi({\bbox k},\omega_n)\phi(-{\bbox
k},-\omega_n)\rangle$, defining the dynamic susceptibility:
\begin{equation}\label{dynsuc}
\chi({\bbox k},\omega_n)=\left(\phi+\omega_n^2+|{\bbox
k}|^\sigma\right)^{-1}.
\end{equation}
From this expression one deduce that the anomalous critical exponent
measuring the divergence of the correlation length at the quantum
critical point is $\eta=2-\sigma$ and the dynamic critical exponent is
$z=\sigma/2$. The denominator of the dynamic susceptibility
(\ref{dynsuc}) has poles in the complex ${\bbox k}$-plane at
$k=\pm\left(-\phi-(2\pi nT)^2\right)^{1/\sigma}$. The closest pole to
the origin is the one corresponding to $n=0$. This pole determines an
exponential decay of the correlation functions. So the correlation
length turns out to be
\begin{equation}\label{correlation}
\xi=\phi^{-1/\sigma}.
\end{equation}
From~(\ref{suc}) and (\ref{correlation}) one deduce the simple ratio
between the critical exponents $\gamma$ and $\nu$ to be equal to
$\sigma$.

To investigate the bulk quantum critical behaviour of the model
(\ref{model}) at zero temperature we transform the sums in
equations~(\ref{scalingpro}) into integrals, i.e.
$$
T\sum_{\omega_n}\to\int\frac{d\omega}{2\pi} \ \ {\rm and} \ \
\frac1V\sum_{\bbox k}\to\int\frac{d^k{\bbox k}}{(2\pi)^d},
$$
to get
\begin{mathletters}\label{bulkt0}
\begin{equation}\label{bulkf}
{\cal F}_0=-\frac{1}{4u_0}(\phi-r_0)^2+\int\frac{d\omega}{4\pi}\int
\frac{d^d{\bbox k}}{(2\pi)^d}\ln\left[\phi+\omega^2
+|\bbox k|^\sigma\right],
\end{equation}
and
\begin{equation}\label{bulkphi}
\phi=r_0+u_0\int\frac{d\omega}{2\pi}\int\frac{d^d{\bbox k}}{(2\pi)^d}
\frac1{\phi+\omega^2+|\bbox k|^\sigma}.
\end{equation}
\end{mathletters}
These equations contains all the necessary information from which we
can extract all what we need about the critical behaviour at zero
temperature. Here we will pay attention to the scaling properties of
the thermodynamic functions in the neighborhood of the quantum
critical point given by
\begin{equation}\label{critical}
r_{0c}=-\frac{u_0}2\int\frac{d^d{\bbox k}}{(2\pi)^d}{|\bbox
k|}^{-\sigma/2}.
\end{equation}
This integral is infrared convergent only for $d>d_<$, where $d_<
=\sigma/2$ defines the lower critical dimension. It is easy to
calculate all the critical exponents for the model we are considering
here in the large $n$ limit by expanding the right hand side of
equation~(\ref{bulkphi}) for small $\phi$. In the expanded expression
one can see a natural emergence of $d_>=3\sigma/2$ as the upper
critical dimension. Here we will give some of these critical
exponents:
\begin{equation}
\gamma=\frac\sigma{d-d_<}, \ \ \
\nu=\frac1{d-d_<}.
\end{equation}

In the remainder of this paper we will investigate the effects of
finite temperature and/or spatial sizes on the bulk zero temperature
critical behaviour.

\section{Finite temperature effects on the quantum critical behaviour}
\label{tempbulk}

At finite temperature we obtain correction terms to the right hand
side of equations (\ref{bulkt0}). These are given by
\begin{mathletters}\label{tcorrection}
\begin{equation}\label{tcorrectionf}
\Delta_{\cal
F}^\tau(T,\phi)=-\frac{k_d}{\sigma\sqrt{\pi}}\Gamma\left(\frac{d}
{\sigma}\right)\phi^{\frac d\sigma+\frac12}\sum_{m=1}^{\infty}
\frac{K_{\frac d\sigma+\frac12}\left(m\frac{\sqrt{\phi}}T\right)}
{\left(m\frac{\sqrt{\phi}}{2T}\right)^{\frac d\sigma+\frac12}}
\end{equation}
for the free energy and
\begin{equation}\label{tcorrectionxi}
\Delta_\xi^\tau(T,\phi)=\frac{2k_d}{\sigma\sqrt{\pi}}\Gamma
\left(\frac{d}{\sigma}\right)\phi^{\frac d\sigma-\frac12}
\sum_{m=1}^{\infty}\frac{K_{\frac d\sigma-\frac12}\left(m\frac
{\sqrt{\phi}}T\right)}{\left(m\frac{\sqrt{\phi}}{2T}\right)^{\frac
d\sigma-\frac12}}
\end{equation}
for the saddle point equation. In expressions (\ref{tcorrection}) we
used the quantity $k_d^{\ -1}=\frac12(4\pi)^{\frac d2}\Gamma(d/2)$ and
$K_\nu(x)$ is the MacDonald function (second modified Bessel
function).
\end{mathletters}

Combining equations (\ref{bulkphi}) and (\ref{tcorrectionxi}) we get,
in the bulk limit corresponding to the geometry $\infty^d\times
L_\tau^z$, the following scaling form (for the saddle point equation)
for dimensions $d$ between the lower $d_<$ and upper critical
dimensions $d_>$
\begin{equation}\label{tzero}
x_\tau\!=\!u_0k_d\frac{y_\tau^{d/z-1}}{2\sigma\sqrt{\pi}}\Gamma\!\!\left(
\frac d\sigma\right)\!\!\!\left[\!\Gamma\!\!\left(\frac12-\frac
d\sigma\right)\!+\!4\!\!\sum_{m=1}^{\infty}\frac{K_{\frac
d\sigma-\frac12}\left(my_\tau\right)}{\left(\frac12my_\tau\right)
^{\frac d\sigma-\frac12}}\right].
\end{equation}
Here the scaling variables are defined by $y_\tau=\sqrt\phi/T$ and
$x_\tau=T^{1-d/z}\left (r_{0c}-r_0\right)$.

For fixed $\sigma$, the solution, $y_0$, of (\ref{tzero}) at the
quantum critical point ($x_\tau=0$) is a universal number. An analytic
solution of this equation cannot be obtained in the general case. Here
we will consider some particular cases:
\begin{equation}\label{sol1}
y_0=\left\{
\begin{array}{ll}
\frac{2\pi}{\sigma}\left(d-d_<\right),
&d-\sigma/2\ll1,\\[.5cm] 2\ln\frac{1+\sqrt{5}}{2}, \ \ \ \ \ \ \ \
&d=\sigma, \\[.5cm]
2\pi\sqrt{\frac{d_>-d}{3\sigma}},
&3\sigma/2-d\ll1.
\end{array}
\right.
\end{equation}

\begin{figure}
\resizebox{5in}{!}{%
\centerline{\includegraphics{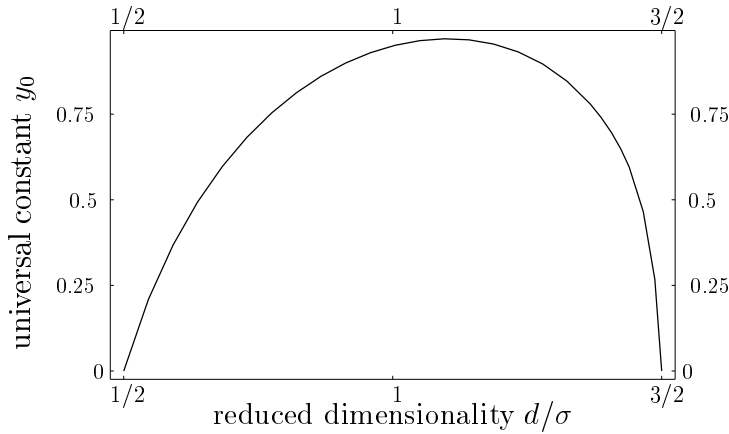}}}
\vspace{.05in}
\caption{Solution of equation~(\protect\ref{tzero}) at $x_\tau=0$
as a function of the reduced space dimensionality of the system}
\label{fig:1}
\end{figure}

At $x_\tau=0$, equation~(\ref{tzero}) can be solved numerically. The
behaviour of the universal constant $y_0$ as a function of the reduced
dimensionality $d/\sigma$ is given in Fig.~\ref{fig:1}. We see that
$y_0$ depend upon the ratio $d/\sigma$ in a universal way for all
values of $\sigma$ smaller than or equal to $2$.

\begin{figure}
\resizebox{5in}{!}{%
\centerline{\includegraphics{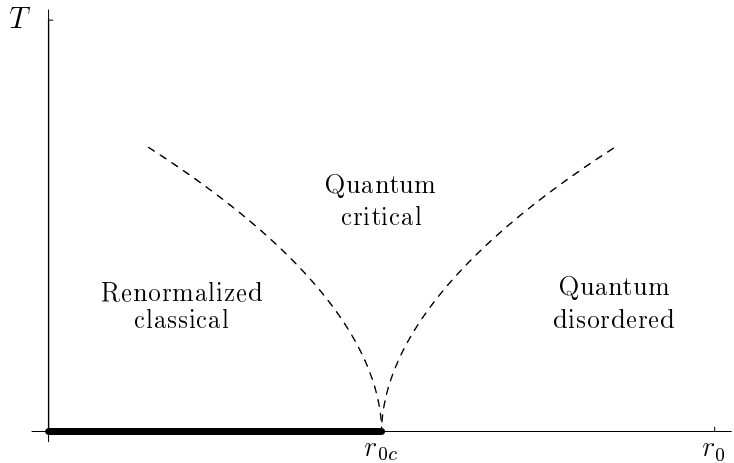}}}
\vspace{.05in}
\caption{Qualitative phase diagram of the model under
consideration in the vicinity of the quantum critical point for
$d=\sigma$. The dashed lines are crossover lines. The tick line shows
the region, where the system is ordered at $T=0$.}
\label{figure2}
\end{figure}

The phase diagram of the model (here we are speaking about the
critical line and crossover lines) in the vicinity of the quantum
critical point is determined by putting $\phi=0$ in the saddle point
equation (\ref{phi}). In the vicinity of the quantum critical point
(\ref{critical}) the phase diagram is determined by
\begin{equation}\label{T0shift}
r_{0c}-r_{0c}(T)=\frac2\sigma k_d T^{\left(2d/\sigma-1\right)}
\Gamma\left(\frac{2d}\sigma-1\right)
\zeta\left(\frac{2d}{\sigma}-1\right),
\end{equation}
where $\zeta(x)$ is the Riemann zeta function. The crossover lines and
the critical line are determined by $|r_{0c}-r_0|\sim T^{1/\nu z}$.
The phase diagram is presented in Fig.~\ref{figure2}. In the different
regions of the phase diagram and for arbitrary dimensions the
correlation length has different behaviors.

For $x_\tau\to\infty$ in the interval $\sigma/2<d<\sigma$, the
correlation length behaves like
\begin{equation}
\xi\sim \left|\frac{T}{r_0-r_{0c}}\right|^{1/(d-\sigma)}.
\end{equation}
This expression shows that, when $T\to0^+$, the correlation length
goes to infinity as quantum effects becomes relevant. In this case
there no quantum phase transition for finite temperature. For the
second interval $\sigma<d<3\sigma/2$ one gets the behaviour
\begin{equation}
\xi\sim \left(\frac{T}{r_0-r_{0c}(T)}\right)^{1/(d-\sigma)}
\end{equation}
for $r_0$ very close but larger than the critical quantum parameter
$r_{0c}(T)$. Here we have a phase transition at finite temperature and
the critical exponents are those of the model in the classical limit.
They are:
\begin{equation}
\gamma=\sigma\left(d-\sigma\right)^{-1}, \ \ \
\nu=\left(d-\sigma\right)^{-1}.
\end{equation}
for $r_0$ less than its critical value the correlation length is
infinite.

For $x_\tau\to-\infty$ the correlation length is temperature
independent. For arbitrary dimension $d_<<d<d_>$ it is given by
\begin{equation}
\xi\sim(r_{0c}-r_0)^{1/(d_<-d)}.
\end{equation}

The particular case $d=\sigma$ is very simple to handle. In this case
equation~(\ref{tzero}) get simpler. Its solution leads to
\begin{equation}\label{dsigma}
\xi^{-z}=T f_\xi\left(x_\tau\right)
\end{equation}
for the inverse correlation length. Here
\begin{equation}
f_\xi\left(x_\tau\right)=2\ {\rm
arcsinh}\left[\frac12\exp\left(-\frac\sigma2
\frac{x_\tau}{k_\sigma}\right)\right]
\end{equation}
is a scaling function, which simplifies in some limiting cases:
\begin{equation}\label{scalingf}
f_\xi\left(x\right)=
\left\{
\begin{array}{ll}
-\frac{\sigma x}{2k_\sigma}                      &x\to-\infty ,   \\[.5cm]
\frac{1}{2}\ln\frac{1+\sqrt{5}}{2} \ \           &x=0,            \\[.5cm]
\exp\left(\frac{\sigma x}{2k_\sigma}\right) \ \  &x\to\infty.
\end{array}
\right.
\end{equation}

From equations~(\ref{scalingf}) one can transparently see the
different behaviours of the correlation length $\xi(T)$ in three
regions:

a) {\it renormalized classical region} with exponential divergence as
$T\to 0$. In this region the system display characteristics of the
ordered ground state. The thermal fluctuations destroys long-range
order at any finite temperature.

b) {\it quantum critical region} with $\xi(T)\sim T^{-2/\sigma}$ and
crossover lines $T\sim|r_{0c}-r_{0}|^{\sigma/2}$. In this region the
system ``notices'' that it is finite in the time direction.

c) {\it quantum disordered region} with temperature independent
correlation length as $T\to 0$. In this region the system has a gap in
the spectrum.

The different regions are qualitatively shown in the phase diagram in
Fig. \ref{figure2}.

\section{Quantum critical finite-size scaling}\label{zerotemperature}
As the temperature is set to zero, equation~(\ref{scalingpro}) turns
into
\begin{mathletters}\label{zerotemp}
\begin{equation}\label{freezero}
{\cal F}_L=\frac{1}{2V}\sum_{\bbox k}\sqrt{\phi+|{\bbox k}|^\sigma}-
\frac{1}{4u_0}(\phi-r_0)^2
\end{equation}
for the free energy and
\begin{equation}\label{saddlezero}
\phi-r_0=\frac{u_0}{2V}\sum_{\bbox k}\frac{1}{\sqrt{\phi+|{\bbox
k}|^\sigma}},
\end{equation}
for the saddle point equation.
\end{mathletters}

Exploring the finite-size scaling in the quantum limit ($T=0$) turns
out to be a difficult task because of the presence of the term
$|{\bbox k}|^\sigma$ in the spectrum of the model. In other words the
problem is how to handle the terms $\sqrt{\phi+|{\bbox k}|^\sigma}$ in
the expressions for the free energy and in that of the saddle point
equation (\ref{zerotemp}). The solution to this problem was given in
Ref.~\cite{chamati94}, where two important identities facilitating the
analysis of the finite-size scaling were presented (see also
Appendix~\ref{appendix1}).

For the $d$-dimensional system with spatial geometry
$L^{d-d'}\times\infty^{d'}$ at zero temperature and periodic boundary
conditions imposed along the ($d-d'$) finite size directions with
linear size $L$ of the system, the corrections to
equations~(\ref{zerotemp}) are given by
\begin{mathletters}\label{asymptoticeq}
\begin{equation}\label{freeasymptotic}
{\cal F}_L=\frac{1}{2}\int\frac{d^d{\bbox
k}}{(2\pi)^d}\sqrt{\phi+|{\bbox k}|^\sigma}-\frac{1}{4u_0}(\phi-r_0)^2
-u_0L^{d+z}\Delta_{\cal F}^L(L^\sigma\phi)
\end{equation}
and
\begin{equation}\label{saddleasymptotic}
\phi=r_0+\frac{u_0}2\int\frac{d^d{\bbox k}}{(2\pi)^d}\frac1{\sqrt{\phi
+|{\bbox k}|^\sigma}}+u_0L^{z-d}\Delta_\xi^L\left(L^\sigma\phi\right),
\end{equation}
where for convenience we introduced the following functions
\begin{equation}
\Delta_{\cal F}^L(y)=\frac{\sigma}{8}\frac1{(4\pi)^{d/2}} {\sum_{\bbox
l}}'\!\!\int_0^\infty\!\!dx\exp\left(\frac{{\bbox l}^2}{4x}\right)
x^{-\frac{\sigma}{4}-\frac d2-1}{\cal
G}_{\frac{\sigma}{2},1-\frac{\sigma}4}
\left(-x^\frac{\sigma}{2}y\right)
\end{equation}
and
\begin{equation}\label{asymptotic}
\Delta_\xi^L(y)=
\frac12\frac1{(4\pi)^{d/2}}
{\sum_{\bbox l}}'\int_0^\infty\!\!dx\exp\left(\frac{{\bbox
l}^2}{4x}\right) x^{\frac{\sigma}{4}-\frac d2-1}{\cal
G}_{\frac{\sigma}{2},\frac{\sigma}4}
\left(-x^\frac{\sigma}{2}y\right).
\end{equation}
Here the primed summation over the vector ${\bbox l}$ is
$(d-d')$-dimensional and the prime indicates that the term
corresponding to ${\bbox l}=0$ is excluded. In the last equation we
used the function~\cite{chamati94}
\begin{equation}
{\cal G}_{\alpha,\beta}(x)=\frac 1{\sqrt{\pi}}\sum_{k=0}^\infty
\frac{\Gamma \left( k+1/2\right)}{\Gamma \left( \alpha k+\beta
\right)}\frac{x^k}{k!}
\end{equation}
Some properties of the functions ${\cal G}_{\alpha,\beta}$ and
$\Delta_\xi^L(y)$ are discussed in Appendices \ref{appendix1} and
\ref{appendix2}, respectively.
\end{mathletters}

\subsection{The finite-size shifted critical quantum parameter}
\label{shifted}
It is well known from finite-size scaling theory that the critical
value of the parameter driving the phase transition is shifted due the
effects of the finite sizes in the system. The aim of this subsection
is to evaluate the distance over which the critical quantum parameter
$r_{0c}$ of the bulk system is shifted. For our concrete model this is
obtained by substituting the parameter $\phi$ by zero. The result is
\begin{equation}\label{shift}
r_{oc}-r_{0c}(L)=\frac{u_0}{2}\frac{\Gamma(d/2-\sigma/4)}{(4\pi)^{d/2}
\Gamma(\sigma/4)}
{\sum_{\bbox l}}'\left(\frac{|{\bbox l}|L}{2}\right)^{-d+\sigma/2}
\end{equation}
and the ($d-d'$) dimensional sum in the right hand side of
equation~(\ref{shift}) is convergent for $d'>d_<$.

In the opposite case $d_<>d'$, the r.h.s of (\ref{shift}) is
divergent. Nevertheless the sum in equation~(\ref{shift}) can be
expressed in terms of the Epstein zeta function, which is a
generalization of the Rieman zeta function. The resulting Epstein
function can be analytically continued beyond its domain of
convergence to give a physical meaning to equation~(\ref{shift}) as
well. In this case the shifted ``pseudocritical'' quantum parameter
$r_{0c}(L)$ corresponds to the center of the rounding of the
singularities of the thermodynamic functions, taking place in the
thermodynamic limit. This point has been investigated in details in
the framework of the finite-size shift of the critical temperature for
the spherical model in Ref. \cite{chamati96}.

Let us also mention that the distance over which the critical quantum
parameter is shifted can be also expressed as
\begin{equation}\label{madshift}
r_{0c}-r_0(L)=\frac{u_0}{2(2\pi)^{\sigma/2}}\frac{\pi^{d'/2}}
{\Gamma(\sigma/4)}{\cal C}_{d,d',\sigma},
\end{equation}
where ${\cal C}_{d,d',\sigma}$ is a the Madelung type constant (cf.
Appendix~\ref{appendix2}).

One can see that the shifted critical quantum parameter $r_{0c}(L)$ is
lower than its bulk bulk critical value $r_{0c}$ for the different
values of $d$, $d'$ and $\sigma$ (which is the ``normal case'', see
Ref. \cite{chamati96}), while the pseudocritical $r_{0c}(L)$ is upper
than the bulk critical quantum parameter. However for the boundary
case when $d'\to d_{<}$ we find that the shift is infinite. This may
be explained with the aid of the behaviour of the Epstein zeta
function at its pole $d'=d_<$. The shift in this case is $\delta
r_0\sim (d'-d_<)^{-1}$ and the appearance of $\mp\infty$ is clear.

\subsection{Finite-size scaling at zero temperature}

In the neighborhood of the quantum critical point, it is possible to
write equations~(\ref{asymptoticeq}) in the scaling forms (for
dimensions between the lower $d_<$ and the upper $d_>$ critical
dimensions)
\begin{mathletters}\label{zeroTscaling}
\begin{equation}\label{zerofscaling}
{\cal F}_L-{\cal F}_0=L^{-d-z}\left[
-D_{d,\sigma}^{\cal F}y_L^{\frac d\sigma +\frac12}
+\frac1{2u_0}x_Ly_L-\Delta_{\cal F}^L\left(y_L\right)\right]
\end{equation}
for the singular part of the free energy and
\begin{equation}\label{zerosscaling}
x_L=\frac{u_0}2D_{d,\sigma}y_L^{\frac d\sigma -\frac 12}
+u_0\Delta_\xi^L\left(y_L\right)
\end{equation}
for the saddle point equation, whose solution is related to the
correlation length. Here the scaling variables are given by
$x_L=(r_{0c}-r_0)L^{d-d_<}$ and $y_L=L^\sigma\phi$, and
\begin{equation}\label{constD}
D_{d,\sigma}=-2D_{d,\sigma}^{\cal F}\left(\frac
d\sigma+\frac12\right)^{-1}=\frac{k_d}{\sigma\sqrt{\pi}}\Gamma
\left(\frac d\sigma \right) \Gamma
\left(\frac 12-\frac d\sigma \right).
\end{equation}
\end{mathletters}

From equations~(\ref{zeroTscaling}) one can see that the singular part
of the free energy and the correlation length are universal scaling
functions of the variable $x_L$ i.e.
\begin{mathletters}\label{amplitude}
\begin{equation}
{\cal F}_s\equiv{\cal F}-{\cal F}_0=L^{-d-z}f_{\cal F}(x_L)
\end{equation}
and
\begin{equation}
\xi=L f_{\xi}(x_L).
\end{equation}
\end{mathletters}
At the quantum critical point i.e. $x_L=0$ the critical amplitudes
$f_{\cal F}(0)$ and $f_\xi(0)$ are depending upon the geometry of the
system and the range of the interaction.

{\it i}) In the region corresponding to $x_L\to\infty$ (in other words
if the quantum parameter $r_0$ is less but very close to its critical
value $r_{0c}$), we obtain
\begin{equation}\label{larger}
\xi=\left(2\frac{r_{0c}-r_0}{u_0D_{d',\sigma}}\right)^{\frac1{d_<-d'}}
L^{\frac{d-d'}{d_<-d'}}.
\end{equation}
This shows how the correlation length tends to infinity as the size of
the system becomes larger. In the bulk system we recover the fact that
the correlation length is infinite in the ordered phase in the large
$n$ limit.

{\it ii}) In the case when the quantum parameter $r_0$ coincides with
its critical value $r_{0c}$, the value of the scaling function
$f_\xi(0)$ determines the critical amplitudes at the critical point.
The value of these critical amplitudes depends upon the dimension $d$
of the system, the number of infinite sizes $d'$ and the range of the
interaction $\sigma$. It can be evaluated analytically in the vicinity
of the borders (determined by the critical dimensions) where the
scaling is valid. The scaling variable vanishes and the function
$\Delta_\xi(y_L)$, in equation~(\ref{zerosscaling}), can be replaced
by its asymptotic form for small argument (\ref{Deltaxi1}). As a
solution for the obtained equation we find
\begin{equation}\label{borders}
\frac{\xi}L=
\left\{
\begin{array}{ll}
\left(\!\frac{2}{(4\pi)^{\frac\sigma4}(d-\frac{\sigma}2)
\Gamma\left(\frac\sigma4\right)
D_{d',\sigma}}\!\right)^{\frac1{\sigma/2-d'}},
\ \ \ &d-\frac\sigma2\ll1,\\[.5cm]
\left(\!\frac{1}{(4\pi)^{\frac{3\sigma}4}(\frac{3\sigma}4-d)
\Gamma\left(\frac{3\sigma}{4}\right)
D_{d',\sigma}}\!\right)^{\frac1{3\sigma/2-d'}},
&\ \ \frac{3\sigma}2-d\ll1.
\end{array}
\right.
\end{equation}

In some particular cases when the function ${\cal G}_{\alpha,\beta}
(x)$ gets simple, namely in the cases $\sigma=1$ and $\sigma=2$ (see
Appendix~\ref{appendix1}), and for some special cases of the
dimensions $d$ and $d'$ it is possible to solve
equation~(\ref{zerosscaling}) numerically. Then one gets
\begin{equation}\label{specialcases}
\frac{\xi}{L}=
\left\{\begin{array}{ll}
0.624798&\rm{for} \ d=1,\ d'=0,\ \sigma=1 \\ [.5cm]
1.511955 &\rm{for}
\ d=2,\ d'=0,\ \sigma=2 \\ [.5cm]
2\ln\left(\frac{1+\sqrt{5}}{2}\right)\ \ &\rm{for} \ d=2,\ d'=1,\
\sigma=2
\end{array}
\right.
\end{equation}

{\it iii}) The last case we consider here is the one corresponding to
the values of the quantum critical parameter smaller that the critical
value i.e. $x_L\to-\infty$. The correlation length is $L$ independent
and it is given by
\begin{equation}\label{smaller}
\xi=\left(\frac{u_0 D_{d,\sigma}}{2\left(
r_{0c}-r_0\right)}\right)^{\frac1{d-d_<}}.
\end{equation}

\section{Interplay between quantum and finite-size effects}
\label{interplay}
In this section we consider the case of a finite system at low
temperatures. In other words we will investigate a system confined in
the finite geometry of the general form
$L^{d-d'}\times\infty^{d'}\times L_\tau^z$ (here we will limit our
discussion only to the case $d'<d_<$). In this geometry, in addition
to the correction terms to equation~(\ref{bulkphi}) given
in~(\ref{tcorrectionxi}) and (\ref{asymptotic}) we have to add also a
correction term which accounts for the combined effects of finite
sizes and finite temperature. This is given by
\begin{mathletters}\label{combined}
\begin{eqnarray}
\Upsilon\left(\phi,L,T\right)&=&\frac{\sqrt{2}L^{-d}}
{\left(2\pi\right)^{\frac{d+1}2}}\sum_m{\sum_{\bbox
l}}'\int_0^\infty\frac{dz}{\sqrt{z}}\exp(-z\phi-\frac{T^2m^2}{4z})
\nonumber\\
& &\times|{\bbox l}|^{-d}\Phi_{d/2-1,\sigma}\left(zL^{-\sigma}|{\bbox
l}|^{-\sigma}\right),
\end{eqnarray}
where
\begin{equation}\label{Phi}
\Phi_{\nu,\sigma}\left(y\right)=\int_0^\infty dx x^{\nu+1} J_\nu(x)
e^{-yx^\sigma}
\end{equation}
was introduced in Ref. \cite{chamati99}. In equation (\ref{Phi}),
$J_\nu(x)$ stands for the Bessel function.
\end{mathletters}

The general equation obtained by putting all together equations
(\ref{bulkphi}), (\ref{tcorrectionxi}), (\ref{asymptotic}) and
(\ref{combined}) can be written in a scaling form whose solution gives
the correlation length as a function depending upon two scaling
variables. This has the general form
\begin{equation}\label{comcorrlen}
\xi=L f_\xi\left(x_L,L T^{1/z}\right)=T^{-1/z} f_\tau\left(x_\tau,L
T^{1/z}\right).
\end{equation}
Actually we see that there will be a competition between the finite
sizes and quantum effects depending upon the quantity $L T^{1/z}$ as
we will see latter. The scaling form (\ref{comcorrlen}) is in
agreement with predictions of Section~\ref{FSSQ}.

The solution of the saddle point equation for a system confined to a
general geometry is very difficult to obtain in an explicit form. Even
in the two limiting cases of ``low-temperature'' ($LT^{1/z}\gg1$) and
``very low-temperature'' ($LT^{1/z}\ll1$) the asymptotics of the
general equation are very complicated. Nevertheless these limits
provide some useful information about the behaviour of the system in
the first or the latter case. Let us mention that the ensuing
mathematical equations simplifies drastically in the case of
short-range forces i.e. $\sigma=2$. In this particular case the
analysis of the saddle point equation is identical to the one
presented in the framework of the quantum rotor spherical model (for
details see Ref. \cite{chamati98}). There is a case when the saddle
point equation takes more simple form. This is namely the particular
case of long-range interaction corresponding to $\sigma=1$. Even in
this case the analysis of the critical behavior is very complicated.

Before doing this let us present the expressions for the asymptotic
forms corresponding to the limiting cases when the finite temperature
effects dominate the finite-size scaling and vice versa.

\subsection{Low-temperature regime $LT^{1/z}\gg1$}
In this regime the finite temperature corrections lead those coming
from the finite size of the system. The scaling form of the saddle
point equation is given by
\begin{mathletters}\label{LTasym}
\begin{eqnarray}\label{LTasymptotic}
x_\tau&=&\frac{u_0}2D_{d,\sigma}y_\tau^{1/z\nu}+u_0\frac{k_d}
{\sigma\sqrt{\pi}}\Gamma\left(\frac d\sigma\right)y_\tau^{1/z\nu}
{\cal K}_2\left(\!\frac
d\sigma\!-\!\frac12\!;\frac{y_\tau}2\right)\nonumber\\ &
&+\frac{2u_0}{(4\pi)^{d/2}}y_\tau^{2(d/\sigma-1)}{\cal
K}_\sigma\left(\frac d2-1;\frac{y_L}2\right),
\end{eqnarray}
where we used the functions
\begin{equation}
{\cal K}_2(\nu;y)=2\sum_{m=1}^\infty(my)^{-\nu}K_{\nu}(2my)
\end{equation}
and
\begin{equation}
{\cal K}_\sigma(\nu;y)={\sum_{\bbox l}}'\int_0^\infty dx
\frac{x^{\nu+1}}{1+x^\sigma} \frac{J_\nu\left(2y{\bbox l}x\right)}
{(y{\bbox l})^\nu}.
\end{equation}
\end{mathletters}
The last function has been proposed in Ref. \cite{singh89}, where the
finite-size scaling properties of the spherical model were discussed.
The analytical properties of this function are considered in the same
reference.

The function ${\cal K}_2(\nu;y)$ introduced by equations
(\ref{LTasym}) is exponentially decreasing for large values of its
argument $y$ (for fixed finite $\nu$). The second function, i.e ${\cal
K}_\sigma(\nu,y)$, is decaying exponentially only in the case of
short-range interaction $\sigma=2$. For the case of long-range
interaction corresponding to $\sigma<2$ its asymptotic form decreases
by a power-law. The two functions are identical in the case
$\sigma=2$.

Let us recall that the temperature enters in the right hand side of
equation (\ref{LTasymptotic}) trough the relation
$y_\tau=\sqrt{\phi}/T$. The last term appearing in the scaling
form~(\ref{LTasymptotic}) is a correction to the bulk system at low
temperature resulting from finite-size effects. The temperature
independent part of this term is nothing but the finite-size
corrections to the bulk system in its classical limit i.e. the case
when the quantum effects are irrelevant. Now, in the case of
short-range interaction ($\sigma=2$) this equation has been analyzed
previously \cite{chamati98}. In this case we get {\it exponential
corrections}. In the case of long-range interactions we have a {\it
power-law corrections}. This seems to be a general characteristic for
systems with long range interaction (for the case of the spherical
model see references \cite{singh89,brankov91}).

In the particular case of the two dimensional system with short-range
interaction $d=\sigma=2$, we get the solution
\begin{equation}
\xi^{-1}\approx \theta T+(2-d')\sqrt{\frac{2\pi}{5\theta}}
\sqrt{\frac TL}\exp\left(-TL\theta\right),
\end{equation}
where $\theta=.962424$ is a universal constant. This universal form is
agrees with the predictions announced in Section \ref{FSSQ}.

Let us quote here the result for the shift of critical quantum
parameter from its bulk value. It is
\begin{eqnarray}\label{shiftTL}
r_{oc}-r_{0c}(L,T)&=&\frac2\sigma u_0k_{d}T^{\left(2d/\sigma-1\right)}
\Gamma\left(\frac{2d}\sigma-1\right)
\zeta\left(\frac{2d}{\sigma}-1\right)\nonumber\\
&&+\frac{u_0}{2T}\frac{\Gamma(d/2-\sigma/2)}{(4\pi)^{d/2}
\Gamma(\sigma/2)}
{\sum_{\bbox l}}'\left(\frac{|{\bbox l}|L}{2}\right)^{-d+\sigma}
\end{eqnarray}
The first term in this equation is the finite temperature shift of the
critical quantum parameter and the second term is nothing but the
finite-size shift in the classical limit (i.e. when the quantum
fluctuations are negligible in the system) divided by the temperature.

\subsection{Very low-temperature regime $LT^{1/z}\ll1$}
Here the finite temperature correction are negligible in comparison
with those coming from the finite-size effects. In this case the
scaling form of the saddle point equation turns into
\begin{eqnarray}\label{Tlasym}
x_L&=&\frac{u_0}2
D_{d,\sigma}y_L^{d/\sigma-1/2}+u_0\Delta_\xi^L\left(y_L\right)\nonumber\\
& &+u_0 y_L^{d'/\sigma-1/2}\frac{k_{d'}}{\sigma\sqrt\pi}\Gamma
\left(\frac{d'}{\sigma}\right){\cal K}_2\left(\frac{d'}\sigma-\frac12;
\frac{y_L}2\right).
\end{eqnarray}

From this equation we can see that the finite temperature corrections
to the finite system at zero temperature are coming mainly from the
$d'$ infinite dimensions. In this case the temperature corrections are
{\it exponential}. In the particular case of two dimensional system
with short-range interaction $d=\sigma=2$ confined to a strip geometry
($d'=1$) at the quantum critical point, we get
\begin{equation}\label{theta}
\xi^{-1}\approx\frac\theta L +\sqrt{\frac{2\pi}{5\theta}}\sqrt{\frac{L}
{T}}\exp\left(-\frac{\theta}{TL}\right).
\end{equation}
In the case when the system is confined to a block geometry ($d'=0$),
we have
\begin{equation}\label{Omega}
\xi^{-1}\approx\frac{\Omega}{L}+\frac1L\left[\frac1{2\Omega}
+\frac\Omega2{\sum_{\bbox l}}'\left(\Omega^2+4\pi^2{\bbox
l}^2\right)^{-3/2}\right]^{-1}\exp\left(-\frac\Omega{LT}\right),
\end{equation}
where $\Omega=1.511955$ is a universal constant. Both results
(\ref{theta}) and (\ref{Omega}) are in agreement with the scaling
predictions of Section \ref{FSSQ}.

The shift of the critical quantum parameter in this case is given by
\begin{eqnarray}\label{shiftLT}
r_{oc}-r_{0c}(L,T)&=&\frac{u_0}{2}\frac{\Gamma(d/2-\sigma/4)}{(4\pi)^{d/2}
\Gamma(\sigma/4)}
{\sum_{\bbox l}}'\left(\frac{|{\bbox
l}|L}{2}\right)^{-d+\sigma/2}\nonumber\\ &&+\frac2\sigma
k_{d'}L^{d-d'}T^{\left(2d'/\sigma-1\right)}
\Gamma\left(\frac{2d'}\sigma-1\right)
\zeta\left(\frac{2d'}{\sigma}-1\right).
\end{eqnarray}

The first term of the right hand side of equation (\ref{shiftLT}) is
discussed in subsection \ref{shifted}. The second term is the
correction to the finite-size shift coming from the temperature. It is
just the finite temperature shift of a $d'$ dimensional system
multiplied by the volume of a $(d-d')$-finite hypercube with linear
size $L$.

\subsection{Short-range case $\sigma=2$}
The equation for the saddle point equation reads
\begin{eqnarray}\label{sigma=2}
r_{0c}-r_0&=&\frac{u_0}{(4\pi)^{\frac{d+1}2}}\Gamma\left(
\frac{1-d}2\right)\phi^{(d-1)/2}\nonumber\\
&&+\frac{\phi^{(d-1)/2}}{(4\pi)^{\frac{d+1}2}}{\sum_{m,{\bbox
l}}}'\frac{K_{(d-1)/2}\left[\phi^{1/2}\left(m^2/T^2+L^2{\bbox
l}^2\right)\right]}{\left[\phi^{1/2}\left(m^2/T^2+L^2{\bbox
l}^2\right)\right]^{(d-1)/2}}.
\end{eqnarray}
This is the simplest form one can get for the saddle point equation.
In this case the dynamic critical exponent $z$, which measures the
anisotropy is $z=1$. The system is symmetric with respect of the
change $L\leftrightarrow T^{-1}$.

The general analysis of equation (\ref{sigma=2}) follows the one
presented in the framework of the quantum rotors model. Here we will
not discuss this point since one can see the details in Ref.
\cite{chamati98}.

\section{Summary}\label{discussion}

The ${\cal O}(n)$ vector $\varphi^4$ model is extensively used in the
analysis of the critical phenomena because of its direct relevance to
the physical reality. In the limit $n\to\infty$ it adds the property
of exact solvabilty at any dimension. This makes it very attractive
for the exploration of the the scaling properties of quantum critical
phenomena as well as finite-size scaling theory.

In the paper we present investigations on the finite-size scaling of
the $\varphi^4$-model in the vicinity of its quantum critical point.
The striking characteristic of the model is the presence of long-range
interaction decaying at large distances $r$ with a power law as
$r^{-d-\sigma}$. We considered the model confined to the general
geometry of the form $L^{d-d'}\times\infty^{d'}\times L_\tau^z$, where
$L$ is the spatial size of the system, $L_\tau\sim T^{-1/z}$ and $z$
is the dynamic critical exponent.

The results are obtained by considering the temperature which governs
the crossover between the classical and the quantum critical behaviors
as an additional temporal dimension.

A detailed investigation of the alteration of the zero temperature
critical behaviour of the model due to the finite temperature is
performed. For dimensions $d_<<d<d_>$, we studied the critical
behaviour of the bulk model in the tree different regions: classical
renormalized, quantum critical and quantum disordered with different
behaviors of the correlation length as function of the temperature.
The behavior of the correlation length in these different regions as a
function of the dimensionality is calculated. Also, some critical
amplitudes are evaluated.

The large $L$ behavior and the scaling forms accounting for the finite
size effects is derived for the free energy and the saddle point
equation. For the finite-size shift of the the critical quantum
parameter, we found $r_{0c}-r_0\sim L^{-1/\nu}$ in accordance with the
general postulates of finite-size scaling. For some particular cases
the behavior of the correlation length and some critical amplitudes
are evaluated.

The study of the general case when the system is confined to the
general geometry $L^{d-d'}\times\infty^{d'} \times L_\tau^z$ i.e. when
the temperature as well as the sizes of the system are finite turns
out to be a very difficult task because of the high anisotropy of the
system due to the parameter $\sigma$. Nevertheless one can make
interesting deductions in some limiting cases: ({\bf i}) in the
low-temperature regime ($LT^{1/z}\gg1$), the finite-size corrections
to the low-temperature behavior are {\it exponentially} small in the
case of short-range interaction and are decreasing with a {\it power
law} in the case of long-range interaction, ({\bf ii}) in the very
low-temperature regime ($LT^{1/z}\ll1$), however, the finite
temperature correction to the finite-size behaviour are always {\it
exponentially} small.

Here, we confined our investigations on the static critical properties
of the model, believing that the spherical limit ($n=\infty$) provides
a useful tool for studying quantum critical phenomena in dimensions
$d>\sigma/2$. The dynamic properties are not well described in this
limit and require loop corrections (see e.g. reference
\cite{sachdev94} in the case of the nonlinear quantum sigma model).

Though derived for the special case of the $\varphi^4$ model with
long-range interaction in the large $n$ limit, the obtained here
results are expected to hold also for many other cases. For example,
recently, a model suitable to handle the joint description of
classical and quantum fluctuations in an exact manner, was considered
in a number of publications~\cite{chamati94,plakida86,tonchev91,%
verbeure92,pisanova93,pisanova95}. This model is a modification of the
$\varphi^4$-lattice model used extensively in the investigation of the
critical behaviour of the the structural phase
transitions~\cite{bruce80}, in the spirit of the self-consistent
phonon approximation method~\cite{plakida86}.

\acknowledgments
This work is supported by The Bulgarian Science Foundation under
Project F608.

\appendix
\section{Some properties of the function ${\cal G}_{\alpha,\beta}
(\lowercase{z})$}
\label{appendix1}
The functions ${\cal G}_{\alpha,\beta}(z)$ were introduced in
Ref.~\cite{chamati94} in order to investigate the zero temperature
finite-size scaling of an anharmonic crystal with long-range
interaction at zero temperature. They are entire function series of
finite order of growth defined by
\begin{equation}\label{gfunction}
{\cal G}_{\alpha ,\beta}(t) =\frac1{\sqrt{\pi}}
\sum^{\infty}_{k=0}\frac{\Gamma\left(k + 1/2\right)}{
\Gamma\left(\alpha k+\beta\right)}\frac{t^{k}}{k!},\ \alpha>0,\ \beta>0.
\end{equation}
One of the most striking properties of this function is that it obeys
the following identity
\begin{equation}\label{identity1}
\frac1{\sqrt{1 + z}} = \int^{\infty}_{0}dx e^{-x}x^{\beta-1}
{\cal G}_{\alpha,\beta} \left(-x^{\alpha}z\right),
\end{equation}
which is obtained by means of term-by-term integration of the series
(\ref{gfunction}). The identity (\ref{identity1}) lies in the basis of
the mathematical investigation of finite-size scaling in quantum
systems with long-range interaction.

In some particular cases the functions ${\cal G}_{\alpha,\beta}(z)$
reduce to known functions. Here we will give some examples which are
of interest for us in this paper:
\begin{mathletters}
\begin{equation}\label{sigma2}
{\cal G}_{1,1/2}(z)=\frac{e^z}{\sqrt{\pi}},
\end{equation}
\begin{equation}
{\cal G}_{1/2,1/4}(-z) = \frac1{4\sqrt{\pi}}
U\left(\frac34,\frac12,z^2\right), \ \ \ \ {\rm for}\ \  z\geq0
\end{equation}
and
\begin{equation}
{\cal G}_{1/2,3/4} (-z) = \frac1{\sqrt{\pi}}
U\left(\frac14,\frac12,z^2\right), \ \ \ \ {\rm for}\ \  z\geq0,
\end{equation}
where $U(a,b,z)$ is the confluent Hypergeometric
function~\cite{abramovitz64}.
\end{mathletters}

If we set in the identity~(\ref{identity1}) $z=y^{-\alpha}$, $y>0$,
and $x=ty$, we will obtain the Laplace transform
\begin{equation}
\frac{y^{\alpha/2-\beta }}{(1+y^{\alpha})^{1/2}}=\int^{\infty}_{0}dx
e^{-yt}t^{\beta-1}{\cal G}_{\alpha,\beta}(-t^{\alpha}),
\end{equation}
from which we derive a new identity by setting $\beta=\alpha /2$:
\begin{equation}
(1+z^{\sigma})^{-1/2}=\int^{\infty}_{0}dx e^{-xz}x^{\sigma/2-1} {\cal
G}_{\sigma,\sigma/2}(-x^{\sigma}).
\end{equation}
With the aid of the last equation we obtained the large $L$ asymptotic
behaviour~(\ref{saddleasymptotic}) from equation~(\ref{saddlezero}).

The integral representation of the functions ${\cal
G}_{\alpha,\beta}(z)$ is obtained with the aid of the Hankel's
integral for the inverse of the gamma function
\begin{equation}\label{hankel}
\frac1{\Gamma(z)}=\frac1{2\pi i}\int_{C}e^u u^{-z}dz,
\end{equation}
where the integration contour $C$ is a loop which starts and ends at
$x=-\infty$ and encircles the origin in the positive sense:
$-\pi\leq\rm{arg}z\leq\pi$ on $C$~\cite{abramovitz64}. This enables to
get the result
\begin{equation}\label{intrep}
{\cal G}_{\alpha,\beta}(z)=\frac1{2\pi i}\int_{C}dv
\frac{e^{v}v^{-\alpha/2+(1-\beta)}}{(v^\alpha-z)^{1/2}}.
\end{equation}
The last result is valid only for $\alpha<1$. The integrand in
(\ref{intrep}) has a branch-point at $v=0$. A more complete and
detailed analysis of the function ${\cal G}_{\alpha,\beta}(z)$ will be
presented in a subsequent paper. This integral representation is
helpful for obtaining the asymptotic behaviour for $z\to\infty$.

In the following we will investigate the asymptotic behaviour of the
functions ${\cal G}_{\alpha,\beta}(-z)$ for real argument $z$. This
may performed by the use of the series
\begin{equation}\label{sqrts}
\frac1{\sqrt{x+t}}=\frac1{\sqrt{2x}}\sum_{k=0}^p\frac{\Gamma(k+1/2)}{k!}
\left(-\frac t x\right)^k+\frac1{\sqrt{2x}}\sum_{k=p}^\infty
\frac{\Gamma(k+1/2)}{k!}\left(-\frac t x\right)^k
\end{equation}
for $x\gg t$ . The second part of the right hand side of the last
equation is nothing but the expansion in series of the hypergeometric
function $_2F_1(a,b;c,x)$ multiplied by some coefficients. So by
simple rearrangement equation (\ref{sqrts}) takes the form
\begin{eqnarray}\label{hyperg}
\frac1{\sqrt{x+t}}&=&\frac1{\sqrt{2x}}\sum_{k=0}^p\frac{\Gamma(k+1/2)}{k!}
\left(-\frac t x\right)^k\nonumber\\
& &+\frac1{\sqrt{\pi x}}\left(-\frac z x\right)^{p+1}
\frac{\Gamma\left(p+3/2\right)}{\Gamma\left(p+2\right)}\
_2F_1\left(1,p+\frac32;p+2,-\frac z x\right).
\end{eqnarray}

Using the integral representation (\ref{intrep}) and the identity
(\ref{hyperg}) we get the asymptotics (for $p\geq1$)
\begin{eqnarray}\label{asymp}
{\cal G}_{\alpha,\beta}(-x)&=&\sum_{k=0}^p(-1)^k
\frac{\Gamma\left(k+1/2\right)}{k!\sqrt{\pi}}
\frac{x^{-k-1/2}}{\Gamma\left(\beta-\alpha(k+1/2)\right)}\nonumber\\
& &+{\cal O}\left(x^{-p-3/2}\right), \ \ \ x\to+\infty.
\end{eqnarray}
In the particular case $\beta=\alpha/2$, the last equation reduces to
\begin{equation}\label{betaonehalf}
{\cal
G}_{\alpha,\alpha/2}\left(-x\right)\simeq-\frac{x^{-3/2}}{2\Gamma(-\alpha)}
\end{equation}

\section{Asymptotics of the function $\Delta_\xi(\lowercase{y})$}
\label{appendix2}
Here we present the asymptotic behaviours of the functions
$\Delta_\xi(y)$, given in (\ref{asymptotic}) for small and large $y$.
These functions are defined by
\begin{eqnarray}\label{Delta_xi}
\Delta_\xi(y)&=&\frac12 (4\pi)^{-d/2}{\sum_{\bbox l}}'\int_0^\infty
dx \exp\left(-\frac{{\bbox l}^2}{4x}\right)\nonumber\\ & &\times
x^{\sigma/4-d/2-1}{\cal G}_{\sigma/2,\sigma/4}
\left(-x^{\sigma/2}y\right).
\end{eqnarray}

With the aid of the Jacobi identity for a $d$-dimensional lattice
sum
\begin{equation}\label{Jacobi}
\sum_{\bbox l}e^{-x{\bbox l}^2}=\left(\frac\pi x\right)^{d/2}
\sum_{\bbox l}e^{-\pi^2{\bbox l}^2/x}
\end{equation}
we transform expression~(\ref{Delta_xi}) into

\begin{eqnarray}\label{Deltaxi}
\Delta_\xi(y)&=&\frac12D_{d',\sigma} y^{d'/\sigma-1/2}\nonumber\\
& &+\frac12\frac{\pi^{\frac{d'}2}}{(2\pi)^{\frac\sigma2}}
\int_0^\infty dx x^{\sigma/4-d'/2-1} {\cal
G}_{\frac\sigma2,\frac\sigma4}
\left(-y\frac{x^{\frac\sigma2}}{(2\pi)^\sigma}\right)
\nonumber\\
&&\times
\left[{\sum_{\bbox l}}'e^{-x{\bbox l}^2}-\left(\frac\pi x
\right)^{d/2}\right].
\end{eqnarray}
In order to be able to get a reasonable expression for the integral in
the last equation we have to avoid the divergence in the square
brackets. To this end we add and subtract from the function ${\cal
G}_{\alpha,\beta}(x)$ its small asymptotic behaviour to the first
order (see Appendix \ref{appendix1}), which enables us to write (after
some algebra)
\begin{mathletters}
\begin{eqnarray}\label{Deltaxi1}
\Delta_\xi(y)&=&\frac12D_{d',\sigma} y^{d'/\sigma-1/2}-
\frac12D_{d,\sigma}y^{d/\sigma-1/2}
+{\cal W}_{d,d',\sigma}(y)\nonumber\\ &
&+\frac{1}{(2\pi)^\sigma}\frac{\pi^{d'/2}} {\Gamma(\sigma/4)}{\cal
C}_{d,d',\sigma},
\end{eqnarray}
where we used the notations
\begin{eqnarray}
{\cal W}_{d,d',\sigma}&=&\frac{\pi^{d'/2}}{(2\pi)^\sigma}{\sum_{\bbox
l}}'
\int_0^\infty dx x^{\sigma/4-d'/2-1}e^{-x{\bbox l}^2}\nonumber\\
& &\times\left[{\cal G}_{\sigma/2,\sigma/4}\left(-\frac{x^{\sigma/2}y}
{(2\pi)^\sigma}\right)-\frac1{\Gamma(\sigma/4)}\right],
\end{eqnarray}
\begin{eqnarray}\label{madelung}
{\cal C}_{d,d',\sigma}&=&{\sum_{\bbox l}}'\int_0^\infty dx
x^{\sigma/4-d'/2
-1}e^{-x{\bbox l}^2}-\pi^{(d-d')/2}\int_0^\infty dx x^{\sigma/4-d'/2-1}\nonumber\\
&=&\lim_{\lambda\to0}\left\{{\sum_{\bbox l}}'\Gamma\left(\frac\sigma4-
\frac{d'}2,\lambda {\bbox l}^2\right)|{\bbox l}|^{d'/2-\sigma/4}
\right.\nonumber\\
&&-\int_{-\infty}^\infty\cdots\int_{-\infty}^\infty\left. dx\Gamma\left(
\frac\sigma4-\frac{d'}2,\lambda {\bbox l}^2\right)|{\bbox l}|^{\frac{d'}2-
\frac\sigma4}\right\}.
\end{eqnarray}
\end{mathletters}
Here $\Gamma(\alpha,x)$ is the incomplete gamma function. The
expression in equation~(\ref{madelung}) is a generalization of the
Madelung type constant. It is $y$ independent. Indeed this equation
defines the finite-size shift of the critical quantum parameter
$r_{0c}$ at zero temperature. It is easy (following
Ref.~\cite{chamati98}) to show that equations (\ref{madelung}) and
(\ref{shift}) are equivalent.

For small $y$ the asymptotic behaviour of $\Delta_\xi(y)$ is given by
the first term in the right hand side of equation~(\ref{Deltaxi1}).

For large $y$ the asymptotics the function $\Delta_\xi(y)$ are
obtained by substituting the large $x$ behaviour of the functions
${\cal G}_{\alpha,\beta}(x)$ from equation~(\ref{asymp}) in the
definition (\ref{Delta_xi}). After some calculations one ends up with
\begin{equation}\label{largeyasy}
\Delta_\xi(y)\simeq -\frac14 y^{-3/2}\frac{(4\pi)^{\sigma/2}}
{\Gamma(-\sigma/2)}\Gamma\left(\frac{d+\sigma}2\right)
\sum_{\bbox l} \left(\frac1{\pi|{\bbox l}|}\right)^{\frac{d+\sigma}2}.
\end{equation}

\end{document}